\documentclass[11pt]{article}
\usepackage{amsfonts,amscd,amsmath,cite,graphicx}

\def\mysection#1{\section{#1}\setcounter{equation}{0}}
\newcommand{\ap}{\alpha_{\text r}}
\newcommand{\e}{\text{e}}
\newcommand{\X}{\mathbf{X}}
\newcommand{\adsd}{\text{AdS}_{d+1}}
\newcommand{\ads}{\text{AdS}_5}
\newcommand{\N}{{\cal N}}
\newcommand{\V}{V_{\text x}}

\newcommand{\tr}{\text{Tr}}

\title{Scaling Violation and Gauge/String Duality}
\author{Oleg Andreev\thanks{Also at Landau Institute for Theoretical Physics, Moscow, Russia. 
Email address: andre@itp.ac.ru}\\
{\it Laboratoire de Physique Th\'eorique de l'\'Ecole Normale Sup\' erieure}\\
{\it 24, Rue Lhomond 75231 Paris Cedex 05, France}}
\date{}
\begin{document}
\maketitle
\begin{abstract}
We explore the possibilities for scaling violation in gauge theories that have string duals.
Like in perturbative QCD, short-distance behaviour yields logarithms that violate the scaling.
\\
PACS number(s): 11.25.Pm, 12.38.Bx  
\end{abstract}
\vspace{-10.5cm}
\begin{flushright}
LPTENS-05-40
\end{flushright}
\vspace{9.5 cm}
\mysection{Introduction}
As is well known, the QCD analysis of large-momentum-transfer processes results in power behavior. In particular, the 
dimensional scaling laws \cite{scaling}
\begin{equation}\label{scaling}
{\cal A}\sim p^{4-n}\left[1+O\left(\frac{1}{p}\right)\right]
\end{equation}
for the asymptotic behavior of fixed-angle scattering hold for a broad range of processes in the physics of hadrons. 
Here, $p$ is a large momentum scale and $n$ is a minimum total number of hadronic constituents (valence quarks). 
In the early days of string theory (dual resonance models), there was an outstanding problem to recover these laws.

Recently, the high-energy behavior of superstring amplitudes was studied in the case of warped spacetime geometries which 
are the products of $\ads$ with some compact five-manifolds \cite{scaling1,scaling2,andreev}. One of the most important results is 
that of Polchinski and Strassler \cite{scaling1}. They proposed a scheme of evaluating high-energy fixed-angle string amplitudes 
and recovered the dimensional scaling laws. Thus, the long-standing problem on the way to a string theory description of hadronic 
processes was solved. 

In fact, what was proposed in \cite{scaling1} is to integrate string tree-level amplitudes in flat four-dimensional space over an additional 
parameter which is nothing else but the fifth coordinate of $\ads$.\footnote{One can also include an integration over coordinates 
of the compact 
five-manifold but it doesn't matter to the scaling laws.} Using this idea, one can readily extend the scheme to string loop amplitudes and, 
moreover, introduce a perturbation series by assuming that perturbation theory in question is a topological expansion. It was 
claimed in \cite{andreev} that as long as the theory is finite,  the amplitudes exactly fall as powers of momentum. Certainly, this 
is not the case in the real world where logarithms violate the scaling. According to QCD, these logarithms are due to short-distance 
processes. For such processes asymptotic freedom makes perturbative series to be valuable at least as far as we are satisfied with 
a few terms in the series \cite{beta-qcd}.  

As follows from above, it is highly desirable to recover the logarithms in a string theory dual. Moreover, the finiteness of superstring loops 
in flat space may be regarded as a consequence of their soft short-distance behavior seen, for example, in the exponential drop of 
fixed angle scattering. Since the short-distance behavior is no longer soft for strings on warped spacetime geometries, it seems plausible to 
raise the issue of the logarithms. 

There are two main obstacles on the way. First, full control of superstring theory on 
curved backgrounds like $\ads$ is beyond our grasp at present. Second, the string theory dual to QCD is unknown. The standard lore 
is that the background metric is given by 
\begin{equation}\label{metric}
ds^2=\frac{R^2}{r^2}\left(f(r)\eta_{\mu\nu}dx^\mu dx^\nu+dr^2\right)+d\Omega^2
\,,
\end{equation}
where $\eta_{\mu\nu}$ is a four-dimensional Minkowski metric. We take $\eta_{\mu\nu}=\text{diag}(-1,1,1,1)$. Since 
$f(r)\approx 1$ in the region of small $r$,  the metric behaves asymptotically as $\ads\times\text{X}$.  

Fortunately, there is also a piece of good news. First, the logarithms of QCD have a short-distance origin. On the string theory side it 
corresponds to small $r$ that is the region where the most important piece of the metric is known. Second, as follows from a 
discussion of \cite{andreev} the nonzero modes of $\mathbf{r}$ and $\boldsymbol{\Omega}$'s do not play a crucial role in the 
derivation of the scaling behavior.\footnote{$\mathbf{X}$, $\mathbf{r}$, and $\boldsymbol{\Omega}$ are taken to be sigma model 
fields on a string worldsheet. $x$, $r$, and $\Omega$ are their  zero modes, respectively.} What turns out to be crucial is a 
warped geometry and a zero mode of $\mathbf{r}$. So, we are bound to learn something if we succeed.

In this paper we address this issue within the simplified model of \cite{scaling1,andreev}. Our aim is to gain some understanding of 
the singular behavior and, as a consequence, scaling violation by doing simple examples. 

Before proceeding to the detailed analysis, we conclude this section by setting the basic framework for computing matrix elements and 
amplitudes within our model.  The part of the worldsheet action which is most appropriate for our purposes is 
simply\footnote{Since we consider only zero modes of $\mathbf{r}$ and $\boldsymbol{\Omega}$, the kinetic terms are due to   
$\mathbf{X}$'s. It is convenient to introduce a function $\ap$ and use the superspace notations of \cite{friedan}. }
\begin{equation}\label{action}
S_0=\frac{1}{4\pi\ap}\int_{\Sigma}d^2zd^2\theta\,\eta_{\mu\nu}\bar D\X^\mu D\X^\nu \,,\quad
\ap=\alpha' \frac{r^2}{R^2}\,.
\end {equation}
Here $\X$ is a two-dimensional superfield and $\Sigma$ is a two-dimensional Riemann surface. We take a constant dilaton and, 
unfortunately, discard a RR background. We will give some remarks on it in conclusion.

As noted above, we assume that perturbation theory at hand is a topological expansion. Schematically, a g-loop amplitude with $n$ 
external legs is of the form
\begin{equation}\label{g-amplitude}
A_n^{(\text{g})}=\int_0^\infty [dr] \int_{{\cal M}_{g.n}} [d\tau] \,F_n^{(\text{g})}
\,,
\end{equation}
where $F_n^{(\text{g})}$ is an integrand of the ''standard'' g-loop string amplitude in Minkowski space 
with $\alpha'$ replaced by $\ap$, $[d\tau]$ is an integration measure for the moduli space ${\cal M}_{g.n}$ of the Riemann 
surface with $n$ punctures, and $[dr]$ is an integration measure for the zero mode of the fifth dimension. Since we follow the line 
of QCD, we will look for divergences of these integrals at $r=0$. However, before starting to discuss examples in 
detail we need to choose a method of regularization to deal with the infinite integrals.  Both the integrals might be divergent, so 
they should be regulated in a consistent way. We will use the method of dimensional regularization that allows us to do so. This 
means that the background to be considered is given by $\adsd\times X$. 

\mysection{Vector Currents}
The simplest example to discuss is a correlator of two vector currents.  We compute it by following the lines of the first-quantized string 
theory. To this end, we build the corresponding vertex operator. Then, we define the correlator as an expectation value of the 
two vertex operators. 

Following \cite{cg}, we take an operator of naive dimension one half $D\X^\mu$ and dress it with an operator of 
naive dimension zero which is a function. In the simplest case, the vertex operator integrated over the worldsheet 
boundary takes the form
\begin{equation}\label{current}
J^\mu(q)=\oint_C dzd\theta\, D\X^\mu \e^{iq\cdot \X}\Psi(\mathbf{r})
\,,
\end{equation}
where $\X$ is a restriction of the superfield on the boundary, $\X^\mu(z,\theta)=
X^\mu(z)+\sqrt{\alpha'}\theta\psi^\mu (z)$, and $q\cdot\X\equiv q_\mu\X^\mu$. $\Psi$ is a solution to the 
linearized Yang-Mills equation on $\adsd$ \footnote{Since we discard the nonzero modes of $\mathbf{r}$ and 
$\boldsymbol{\Omega}$'s, we set the corresponding YM connections to be zero. However, in such an approximation there is gauge 
invariance $A_\mu\rightarrow A_\mu+\partial_\mu\Lambda$. It is fixed by $\partial\cdot A=0$.}
\begin{equation}\label{DF}
\bigl[r^2\partial_r^2-(d-3)r\partial_r-q\cdot q\,r^2\bigr]\Psi(r)=0
\,.
\end{equation}
For a solution to be nonzero at $r=0$ for $d=4$, we choose
\begin{equation}\label{Psi}
\Psi(qr)=q^\nu r^\nu K_\nu (qr)\,,\quad \nu=\frac{d}{2}-1
\,,
\end{equation}
where $q=\sqrt{q\cdot q}$.

A couple of comments are in order:

\noindent (1) Since the current is conserved, it obeys $q\cdot J=0$. In the approximation we are using this is obvious. Indeed, 
\begin{equation*}
q\cdot J=-i\Psi(qr)\oint_C dzd\theta\, D\,\e^{iq\cdot \X}=0
\,
\end{equation*}
as a total derivative. This is the reason why we consider the worldsheet with boundaries or, equivalently, a spacefilling brane.

\noindent(2) There is a subtle point. The use of this approximation is legitimate only for $q^2=0$.  Certainly, this is not what we need. 
However, it seems that it is safe to go off shell, at least  for rather small $q^2$, as it follows from stringy calculations of the 
renormalization constants \cite{beta}. We will return to this issue below.

Now that we have the vertex operators for the currents, we can focus on the correlator. On general grounds, it takes 
the form 
\begin{equation}\label{2currents0}
i\int \frac{d^dk}{\left(2\pi\right)^d}\,\langle J^\mu(q)\,J^\nu(k)\rangle =\left(q^\mu q^\nu-\eta^{\mu\nu}q^2\right)
\tr (\lambda_1\lambda_2)\,\Pi(q^2)
\,,
\end{equation}
where $\Pi(q^2)$ is given by a perturbative series. Since we assume that the {\it flavor} symmetry group is $U(N_f)$, 
we include a Chan-Paton factor $\tr (\lambda_1\lambda_2)$. 

It seems natural to try a unit disk (upper half plane) as the worldsheet to leading order. We take a covariant measure 
for the zero modes $\sqrt{-g}d^{d+6}\xi$. As to the nonzero modes of $\mathbf{X}$'s, they are quantized in an ordinary way as follows from the 
worldsheet action \eqref{action}\footnote{We use the point splitting method to regulate the propagators \cite{sw}.}
\begin{equation}\label{props}
\langle \mathbf{X}^\mu (z_1,\theta_1)
\mathbf{X}^\nu(z_2,\theta_2)\rangle =\ap\eta^{\mu\nu}
\bigl(G(z_{12})-\theta_1\theta_2 K(z_{12})\bigr)
\,,
\quad
z_{12}=z_1-z_2\,.
\end{equation}
It is clear that the integration over the zero modes of $\mathbf{X}$'s produces a delta-function, while that of 
$\boldsymbol{\Omega}$ 's gives a volume of the compact space X. The integration over the nonzero modes of $\mathbf{X}$'s 
also does not require much work, so we find
\begin{equation}\label{2currents}
\Pi^{\left(0\right)}(q^2)=\frac{1}{2}\,\N^{\,2}\V\,
\mu^{4-d} 
\int_0^1 dz
\left[\bigl(\partial_{z}G(z)\bigr)^2-K^2(z)\right]
\int_0^\infty dr\,\left(\frac{R}{r}\right)^{d+1}\,\Psi^2(qr)\, \ap^2\,
\e^{\,\ap q^2G(z)}
 \,,
\end{equation} 
where $\N$ is a normalization factor which will be fixed shortly. As usual in dimensional regularization, we have introduced an 
arbitrary scale $\mu$ to account for the right dimension. The translational invariance is fixed by setting the second vertex operator 
at the origin. As a consequence, a factor $\frac{1}{2}$ is accounted to the right hand side of \eqref{2currents}. 

Using the integral representation 
\begin{equation}\label{int}
K_\nu(\eta)=\frac{1}{2}\eta^{-\nu}\int^{\infty}_0 d\rho\,\frac{\e^{-\frac{1}{2}(\eta^2\rho+\frac{1}{\rho})}}
{\rho^{\,\nu+1}}
\end{equation}
and keeping only the singular term and a $q$-dependent piece of the finite term, we find at $d=4-2\varepsilon$
\begin{equation}\label{int1}
\mu^{4-d}\int_0^\infty dr\,r^{3-d}\,\Psi^2(qr)\,\e^{\,\ap q^2 G(z)}=\frac{1}{2\varepsilon}+\frac{1}{2}\ln\frac{\mu^2}{q^2}+
\dots
\,.
\end{equation}

We are now in a position to perform the integral over $z$. It is trivial as one can see by substituting the explicit expressions for the 
propagators 
\begin{equation}\label{diskprop}
G(z_{12})=-2\ln\left(4\sin\pi z_{12}\right)
\,,\quad
K(z_{12})=\frac{2\pi}{\sin\pi z_{12}}\,.
\end{equation}

Finally, we use minimal subtraction to get rid  of the pole term $\frac{1}{\varepsilon}$. Thus
\begin{equation}\label{2currents1}
\Pi^{\left(0\right)}(q^2)=
-\frac{1}{4\pi^2}\ln\frac{q^2}{\mu^2}
\,.
\end{equation}
We have fixed the normalization by setting $\N=\frac{1}{2\pi^2\alpha'}\bigl(R\V\bigr)^{-\frac{1}{2}}$. Although this looks like the 
desired QCD result \cite{shifman}, one more step is needed.\footnote{Of course, the number of colors is missing. We will return to this 
issue in Sec.4.} On the right hand side of \eqref{2currents1} $q$ is small, while we are 
looking for the asymptotic behaviour for large $q$. What saves the day is the known fact from the renormalization of the vacuum 
polarization that the leading logarithmic behavior for $q\rightarrow\infty$ is related with the pole $\frac{1}{\varepsilon}$ exactly 
as in \eqref{int1}. Thus, Eq.\eqref{2currents1} holds for large $q$. 

At this point, a few remarks are in order.

 (1) Rather than using dimensional regularization, we could use a regularization scheme with a short distance cutoff. 
It is very easy to see a relation between $\varepsilon$ and the cutoff. Evaluating the integral \eqref{int1} near $r=0$, 
we find $\int_{a}^{l}\frac {dr}{r}=\frac{1}{2}\ln\frac{l^2}{a^2}$ at $d=4$. Thus, $\frac{1}{\varepsilon} =\ln\frac{l^2}{a^2}$. 

(2) In perturbative QCD the correlator is given by a series of Feynman diagrams. A dominant diagram looks like 
that in Fig.1. 

%
\vspace{.6cm}
\begin{figure}[ht]
\begin{center}
\includegraphics{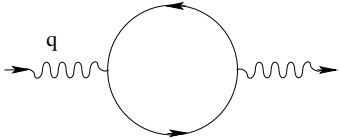}
\caption{\small{A Feynman diagram for the vector current two-point correlator (vacuum polarization). Solid and wavy 
lines denote fermions and gauge bosons, respectively.}}
\end{center}
\end{figure}
\noindent The evaluation of the diagram results in an expression of the form
\begin{equation}\label{qcd}
\Pi(q^2)\sim\mu^{4-d} \int_0^1 d\alpha\,
\alpha\left(1-\alpha\right)
\int_0^\infty dt\, t^{1-\frac{d}{2}}\,
\e^{-t\left(m^2+\alpha\left(1-\alpha\right)q^2\right)}
 \,,
\end{equation} 
where $m$ is a fermion mass. $t$ and $\alpha$ are the Schwinger and Feynman parameters, respectively. 

Let us now compare Eqs. \eqref{2currents} and \eqref{qcd}. Although there is some similarity, these formulae are not 
the same. Nevertheless, we can reach an agreement if we set $m^2=0$ in \eqref{qcd} and restrict ourselves to the singular 
and $q$-dependent pieces. Then up to numerical factors both the integrals are reduced to
\begin{equation*}
\Gamma\Bigl(2-\frac{d}{2}\Bigr)\left(\frac{q^2}{\mu^2}\right)^{\frac{d}{2}-2}
\,.
\end{equation*} 
Equivalently, we could look for the singularities at the lower limits and reach an agreement by identifying the variables $r^2=t$ and 
$z=\alpha$. We discard momentum independent constants because they are irrelevant for large $q$.

(3) At present, it is not clear whether the representation we are using can be taken literally but it seems to us quite safe. Let us mention 
that in the literature a curious way of converting loop diagrams  into trees was discussed a long time ago \cite{barv}. This line of 
thought has recently attracted some attention in the context of the AdS/CFT correspondence. In particular, it was argued that these 
trees are nothing else but tree diagrams in AdS space \cite{gopa}. From this point of view it seems that there is no a paradox: we 
represent the one-loop diagram of QCD as the tree level diagram of string theory in AdS background.\footnote{From the viewpoint 
of string theory with a constant tension it looks like a resummation of the perturbation series. This was also observed by 
investigating the Regge limit \cite{andreev1}.} 

(4) It is worth mentioning that the asymptotic behavior \eqref{2currents1} has been derived within an effective 5-dimensional 
Yang-Mills theory in $\ads$ space \cite{son}. Unlike that, our approach admits a straightforward extension to higher orders of  
perturbation theory. 

Let us now see how it works in the next-to-leading order. Again, our aim is to derive the leading asymptotics. To this end, we take a 
cylinder $C_2$ as the worldsheet. We describe $C_2$ as the region 
\begin{equation}\label{cylinder}
0\leq \Re z \leq 1\,,
\quad
z\equiv z+i\tau
\end{equation}
on the complex plane whose metric is $ds^2=dzd\bar z$.

Since we are interested in the case of $\adsd$, the problem is to extend the modular measure of open string to $d$ flat
dimensions. This was much studied to compute perturbative field theory amplitudes via string theory in the 
$\alpha'\rightarrow 0$ limit (see, e.g., \cite{beta}). In making our further analysis, we will adopt the proposal 
of \cite{joebook} according to which the interpretation of each factor is transparent as it follows from the point-particle results. 
The computation of the correlator is very similar to what we have already done at the tree level. Thus, we obtain\footnote{Note that 
it vanishes at $d=10$. For a discussion, see, e.g., \cite{propagators}.}
\begin{equation}\label{currents-1}
\begin{split}
\Pi^{(1)}(q^2)&=
\frac{i}{2}\N^{\,2}\,\V
\,g^2 \,N_f
\mu^{8-2d}
\int_0^\infty dr\,\biggl(\frac{R}{r}\biggr)^{d+1}\,\Psi^2(qr)\, \ap^{4-\frac{d}{2}}
\int_0^\infty d\tau\,\tau^{-\frac{d}{2}}\left[\eta(i\tau)\right]^{2-d} \\
&\quad\times
\sum_{(\alpha,\beta)\neq(1,1)}
(-1)^{\alpha^2+\beta^2}
\left[\frac{\theta_{\alpha\beta}(0,i\tau)}{\eta(i\tau)}\right]^{\frac{d-2}{2}}
\int_0^{i\tau}  dz
\left[\left(\partial_{z}G(z)\right)^2-K^2_{\alpha\beta}(z)\right]
\e^{\,\ap q^2G(z)}
 \,,
\end{split}
\end{equation} 
where $\eta$ is the Dedekind eta function. We exclude the sector $(1,1)$ from the sum over the spin structures, so 
the fermions $\psi$'s have no zero modes. Again the translational invariance is fixed by setting the second vertex operator at the 
origin and taking into account a factor $\frac{i\tau}{2}$. Since the worldsheet is the cylinder, we need to insert a relative factor $g^2$. 
As usual in string computations using dimensional regularization \cite{beta}, $g^2$ is accompanied with 
$\left(\alpha'\mu^2\right)^{2-\frac{d}{2}}$ that in the problem at hand becomes $\left(\ap\mu^2\right)^{2-\frac{d}{2}}$.

As discussed in introduction, we are interested in the region of small $r$ or, equivalently, large momenta. As a consequence, 
$\ap$ is small. Thus, the integral over $\tau$ may be studied along the lines of the old days $\alpha'\rightarrow 0$ limit  
which is now $\ap\rightarrow 0$. It is well known that in this case only the neighbourhood of $\tau=\infty$ contributes 
to divergences. At this point, it is worth saying that this is {\it not} the supergravity approximation. $\alpha'/R^2$ may be 
large but what is really small is $r^2\alpha'/R^2$. 

To proceed further, we set both the vertex operators on a single boundary $\Re z=0$. A little algebra shows that the sector $(1,0)$ 
dominates for large $\tau$. At this point, it is useful to define new variables
\begin{equation}\label{new}
\varphi=-iz/\tau
\,,\quad
t=2\pi\ap\tau
\,.
\end{equation}
Taking the leading asymptotics for the propagators
\begin{equation*}\label{props1}
G(\varphi_{12})=2\pi\tau\left(\varphi_{12}^2-\varphi_{12}\right)
\,,\quad
K_{10}(\varphi_{12})=2\pi\tau\,,
\end{equation*}
and the theta functions, we get
\begin{equation}\label{currents-2}
\Pi^{(1)}(q^2)=-
\left(\frac{4}{\pi}\right)^{\frac{d}{2}}
g^2\,N_f\,
\left(\frac{\pi R}{\mu^2}\right)^{d-4}
\int_0^1 d\varphi
\left(\varphi^2-\varphi\right)
\int_0^\infty \frac{dt}{t}t^{2-\frac{d}{2}}\e^{\,t\,q^2\left(\varphi^2-\varphi\right)}
\int_0^\infty dr\,r^{3-d}\,\Psi^2(qr) 
 \,.
\end{equation} 

The result has the factorization property. It is not so surprising since we can anticipate from the QCD side that the leading asymptotics 
is given by a diagram shown in Fig.2. Indeed, the first factor is  what QCD analysis provides in the case of massless 
fermions \eqref{qcd}, while the second is not exactly the left hand side of \eqref{int1}. The difference is due to the exponential 
$\e^{\,\ap q^2 G(z)}$. Fortunately, it is irrelevant for the region of small $r$ and, moreover, for the terms of interest 
as those on the right hand side of Eq.\eqref{int1}. Note also that the integral converges for large $r$ because of the 
exponential falloff of $\Psi(qr)$. 

%
\begin{figure}[ht]
\begin{center}
\includegraphics{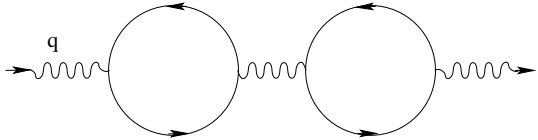}
\caption{{\small A reducible diagram dominating in the next-to-leading order.}}
\end{center}
\end{figure}

It is straightforward to find the double pole term $\frac{1}{\varepsilon^2}$ at $d=4-2\varepsilon$. The evaluation of the first factor 
results in $\frac{1}{6}\left(-\frac{1}{\varepsilon}+\ln\frac{q^2}{\mu^2}\right)$, while that of the second -
$\frac{1}{2}\left(\frac{1}{\varepsilon}-\ln\frac{q^2}{\mu^2}\right)$. Again, we keep only the singular terms and $q$-dependent pieces 
of the finite terms. It is well-known that the renormalization program 
is formulated for one-particle irreducible diagrams that in the case of interest means that we have to remove the poles in each 
term. The use of minimal subtraction yields
 \begin{equation}\label{currents-3}
\Pi^{(1)}(q^2)
=\frac{2}{3}\,g^2N_f
\ln^2\frac{q^2}{\mu^2}
\,.
\end{equation} 
By the same argument we used in the discussion of $\Pi^{(0)}(q^2)$, the result holds for large $q$.

\mysection{Light Mesons}
Perhaps more significant theories on $\ads$ can be applied with some success to the physics of mesons \cite{son,stan}. This motivates 
us to consider the vector mesons. 

We proceed, as before, by first building the corresponding vertex operator. In fact, we have already done much work by doing so for the 
vector current. The discussion differs in two respects:

\noindent (1) We introduce a polarization vector $\xi_\mu(p)$. As a result, the vertex operator \eqref{current} takes the form
\begin{equation}\label{vector}
{\cal O}(p,\xi)=\oint dzd\theta\, \xi\cdot D\X\,\e^{ip\cdot \X}\Psi(\mathbf{r})
\,.
\end{equation}
In the approximation we are using it is invariant under $\xi_\mu\rightarrow\xi_\mu +\phi (p)p_\mu$ as such a shift leads to a 
total derivative. We fix this gauge degree of freedom by setting $\xi\cdot p=0$;

\noindent (2) We choose a solution to Eq.\eqref{DF} with $q\cdot q=-m^2$ such that it vanishes at $r=0$ for $d=4$. Thus
 \begin{equation}\label{Psi-meson}
\Psi(mr)=m^\nu r^\nu J_\nu (mr)\,,\quad \nu=\frac{d}{2}-1
\,,
\end{equation}
with $m$ being a meson mass. Here we should emphasize a subtle point. The use of the approximation is legitimate only for $m^2=0$ 
unless one uses a consistent prescription to go off shell, e.g., as in \cite{beta}. It seems safe at least for small deviations. We will therefore 
consider the light mesons.

Let us now turn to a matrix element of the two meson operators. As in the preceding example, we define it as an expectation value of 
the vertex operators. We have
\begin{equation}\label{xi}
\langle {\cal O}(p_1,\xi_1){\cal O}(p_2,\xi_2)\rangle = \tr (\lambda_1\lambda_2)
\bigl(\xi_1\cdot\xi_2 \,F+\dots\bigr)\delta^{(d+1)}(p_1+p_2)
\,,
\end{equation}
where $F$ is given by a perturbation series. The three dots represent other terms that are higher order in $m^2$ and hence do not make 
a significant contribution as long as we consider the light mesons. 

It seems plausible to try a unit disk as the worldsheet to leading order. Then $F^{(0)}$ can be read from Eqs. \eqref{2currents0},
\eqref{2currents}, and \eqref{diskprop}. We find
\begin{equation}\label{2mesons}
F^{(0)}=1 
\,.
\end{equation} 
It is important to remember that there is a great difference between the current and meson vertex operators. For the mesons the 
integral over $r$ is convergent that allows us to include it into the corresponding normalization factor. Thus we have set 
\begin{equation*}
\N=\frac{1}{2\pi\alpha' m}\left(2^d\pi^{d+1}\V \int_0^\infty dr\,\left(\frac{R}{r}\right)^{d-3}\,\Psi^2(mr)\right)^{-\frac{1}{2}}
\,.
\end{equation*}

We can go further and consider the next-to-leading order. In doing so, we extend the formalism of the previous section to the case of interest. 
We take the cylinder $C_2$ as the worldsheet. Then the correction can be seen directly from Eq. \eqref{currents-2}. Keeping only the pole 
term $\frac{1}{\varepsilon}$ results in
\begin{equation}\label{2mesons1}
F^{(1)}=\frac{8}{3}\,g^2N_f\,\frac{1}{\varepsilon}
\,.
\end{equation} 
This must be renormalized. One possible way to deal with the problem at hand is to renormalize the operator ${\cal O}(p)$. Again, we choose 
minimal subtraction. This gives
\begin{equation}\label{ren}
{\cal O}_0=\sqrt{Z}{\cal O}\,,
\quad
Z=1-\frac{8}{3}\,g^2N_f\,\frac{1}{\varepsilon}
\,,
\end{equation}
with ${\cal O}_0$ being the bare operator. 

To complete this part of the story, it is worth writing down the anomalous dimension of ${\cal O}$.  Since it is simply related to 
the residue in \eqref{ren}, we find 
\begin{equation}\label{dim}
\gamma_{\cal O}=\frac{8}{3}\,g^2N_f
\,.
\end{equation}

Even though the approximation we used is not very good, we can still benefit from it. The point is that the vertex operators \eqref{vector} 
are gauge invariant. As in gauge theory, this makes getting the corresponding $\beta$-function easier. As an example, consider 
elastic scattering of mesons. The amplitude is defined as an expectation value of the four vertex operators
\begin{equation}\label{ampl}
\langle {\cal O}(p_1,\xi_1)\dots {\cal O}(p_4,\xi_4)\rangle=
{\cal A}_4\delta^{(4)}(p_1+\dots +p_4)
\,.
\end{equation}

To keep things as simple as possible, we specialize to  a convenient structure in the kinematic 
factor.\footnote{See, e.g., \cite{schwarz}.} It is $\xi_1\cdot\xi_3\,\xi_2\cdot\xi_4$. Thus, we have 
\begin{equation}\label{amp1}
{\cal A}_4=\tr (\lambda_1\lambda_2\lambda_3\lambda_4)\bigl(\xi_1\cdot\xi_3\,\xi_2\cdot\xi_4\, {\cal A}+
\dots\bigr)
\,,
\end{equation}
where ${\cal A}$ is given by a perturbation series. For the mesons the integral over $r$ is finite. It means that ${\cal A}^{(0)}$ is finite. 
On the other hand, ${\cal A}^{(1)}$ is divergent as the integral over $\tau$ diverges. Now, because of gauge invariance, it immediately 
comes to mind to relate the divergent part of ${\cal A}^{(1)}$ with ${\cal A}^{(0)}$ in the same manner as above. Keeping only the 
pole term $\frac{1}{\varepsilon}$ gives
\begin{equation}\label{div}
{\cal A}^{(1)}=\frac{8}{3}\,g^2N_f\,\frac{1}{\varepsilon}{\cal A}^{(0)}
\,.
\end{equation}

Assuming that ${\cal A}^{(0)}$ is proportional to $g^2$ and involving minimal subtraction, we are in a position to write down 
the $\beta$-function. It is given by
\begin{equation}\label{beta}
\beta(g)=\frac{8}{3}\,g^3N_f
\,.
\end{equation}

In fact, we have been a little cavalier here. It is worth noting that the problem of interest differs from renormalization of the 
Yang-Mills theory in two respects:

\noindent (1) In general,  ${\cal A}^{(0)}$  grows more rapidly than $ g^2$. For example, it may go like $g^4$ as it is clear from a Born 
diagram shown in Fig.3.
%
\vspace{.6cm}
\begin{figure}[ht]
\begin{center}
\includegraphics{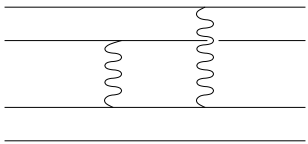}
\caption{{\small A typical Born diagram for pion elastic scattering.}}
\end{center}
\end{figure}
The bare coupling is now defined by $g_0=\frac{g}{\sqrt[4] Z}$. This leads to a correction factor $\tfrac{1}{2}$ in the expression 
\eqref{beta}. 

\noindent (2) There is some combinatorics that also impacts the numerical prefactor of the $\beta$-function. In the example we are 
considering the two diagrams shown in Fig. 4 are equivalent. This increases the prefactor via the corresponding increase of the 
residue.
%
\vspace{.6cm}
\begin{figure}[ht]
\begin{center}
\includegraphics{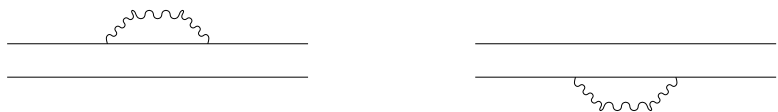}
\caption{{\small Two possibilities for virtual gluons.}}
\end{center}
\end{figure}
We will not dig deeper into these issues. Our reasoning is that they impact the absolute value of the prefactor rather than 
its sign. So, the bottom line is that the model is not asymptomatically free. 

A final remark: having derived the anomalous dimension and $\beta$-function, it seems to be time to involve the Callan-Simanzik 
equation to resum logarithms. We will not do so because there is no asymptotic freedom.

\mysection{Concluding Comments}

There is a large number of open problems associated with the circle ideas exploited in this paper. In this section we list a few.

Here we used the simplified model without any RR background as we still lack the description of such a background within the NSR 
formalism. In particular, the solution of string theory on $\ads$ or on its nonconformal deformation that 
could help us is unknown. So, it is not a big surprise that the number of colors $N_c$ is missing in all the expressions we obtained. 
From this point of view we are in a situation very similar to that in the Skyrme model when the model describes color singlets. 
To include $N_c$, one has to add an extra term into the action. For example, this was done by adding the Wess-Zumino term 
into the original Skyrme action \cite{ed}. The common wisdom is that the worldsheet theory with a warped background 
metric is conformal by virtue of a proper RR background. It seems that this is not the whole story. Our hope is that it might also help 
to get a right correction to the $\beta$-function (see Eq.\eqref{beta}) and, as a consequence, asymptotic freedom. 

A related problem is to understand the nonzero modes of $\mathbf{r}$ and even the transverse fields $\boldsymbol{\Omega}$. 
It is clear that this would allow us to define vertex operators for arbitrary large $q^2$ and avoid the problem with off shell 
continuation. On the other hand, there are some claims that the nonzero modes of $\mathbf{r}$ and even those of $\mathbf{X}$'s 
might be effectively irrelevant in the worldsheet path integral \cite{gopa}. What really happens still remains to be seen.

In this paper we pursued the line of thought borrowed from QCD. Another possible way is to deform the background metric 
for {\it small} $r$ or even somehow cut AdS space to get the desired scaling violation.\footnote{Note that for the 
AdS geometry truncated at some large value of $r$ a rough estimate of \cite{andreev} gives an exponential correction which 
violates the scaling already to leading order.} It would be interesting to see whether the QCD line of thought is equivalent to that or not. 

Finally, one of the possible interpretations for the models of \cite{andsiegel} is that they may be thought of as string theory in  
spacetime whose fifth coordinate is latticized. If one tries to do some computations along the lines of Section 2, this  will 
require the use of lattice regularization for consistency. It suggests that for the models to be well defined, all the coordinates $X$ 
must be latticized after the Wick rotation to Euclidean spacetime. We believe that this issue is worthy of future study.

\vspace{.5cm}
\noindent{\bf Acknowledgments}

\vspace{.25cm}
We would like to thank C. Bachas for useful and stimulating discussions concerning this subject and S. Brodsky, G.F. de Teramond, and 
A.A. Tseytlin for comments on the manuscript. We would also like to acknowledge the hospitality of the Yukawa Institute, 
where a portion of this work was completed. The work is supported in part by CNRS and Russian Basic Research Foundation Grant 
05-02-16486.
\vspace{.25cm}



\end{document}